# Representation learning with reward prediction errors

## William H. Alexander[1,2,3] * | Samuel J. Gershman[4,5]

[1]Center for Complex Systems & Brain Sciences, Florida Atlantic University

[2]Department of Psychology, Florida Atlantic University

[3]The Brain Institute, Florida Atlantic University

[4]Center for Brain Science, Harvard University

[5]Department of Psychology, Harvard University

**Correspondence**
William H. Alexander
Email: walexander@fau.edu

**Funding information**
This material is based upon work supported by the Air Force Office of Scientific Research under award number FA9550-20-1-0413.

The Reward Prediction Error hypothesis proposes that phasic activity in the midbrain dopaminergic system reflects prediction errors needed for reinforcement learning. Besides reward processing, dopamine is implicated in a variety of functions without a clear relationship to reward prediction error. Dopamine levels influence perception of time, dopamine bursts precede motor response, and the dopamine system innervates regions of the brain, including hippocampus and prefrontal cortex, whose function is not specific to reward. We propose a common theme linking these functions is representation, and that dopamine prediction errors, in addition to driving associative learning, can also support the acquisition of adaptive state representations. In a series of simulations, we show how this extension can account for the role of dopamine in temporal and spatial representation, motor response, and abstract categorization tasks. By extending the role of dopamine signals to learning state representations, we resolve a critical challenge to the Reward Prediction Error hypothesis of dopamine function.

# 1 | INTRODUCTION

The dopaminergic (DA) system is implicated in a wide variety of functions, including working memory[1], motor control[2], time perception[3], and value-based learning[4]. Given the range of functions to which DA contributes, it is unsurprising that the neuroætiology of clinical disorders, including schizophrenia[5] and depression[6], as well as diseases involving motor impairment such as Parkinson's Disease (PD)[7], points to dysfunction of the DA system. Drugs of abuse, including cocaine[8] and methamphetamine[9], target the DA system, and their addictive potential is presumed to relate to their ability to disrupt normal DA function[10].

Considering the far-reaching effects of DA function and dysfunction, considerable effort has been made to characterize the underlying mechanisms by which DA contributes to behavior and learning. Over the past 30 years, the Reward Prediction Error (RPE) hypothesis has become perhaps the single-most dominant account of midbrain DA function[11]. Under computational models of reinforcement learning (RL), RPEs drive associative learning between rewards and stimuli that predict them[12]. While the RPE hypothesis is supported by hundreds of empirical studies of DA function and has proven useful in interpreting and predicting behavior and brain activity, it is unclear whether and how RPE signals may contribute to other functions in which DA is implicated.

Apart from its proposed role in learning, DA is implicated in processes not directly tied to reinforcement. While the influence of DA on time perception, including optogenetic manipulation of RPE-like DA activity[13], is well- established [3,14,15], it is presently unclear what role putative DA RPE signals may play in temporal representation. Increased DA levels tend to "speed up" the internal clock, resulting in consistent overestimation of elapsed time intervals, while decreased DA levels slow the internal clock[16] – indeed, unmedicated individuals with PD consistently underestimate elapsed time[17]. However, while RL models generally incorporate a mechanism for tracking time intervals, the temporal representation underlying such models generally serves as the substrate on which RL occurs, and its dynamics are not the target of RL itself[12].

The DA system interacts extensively with regions implicated in spatial representation and navigation, especially hippocampus and surrounding areas[18–20], where it appears to be involved in the formation[21] and stability[22] of memories, including memory for spatial locations[23]. Place cells in hippocampus appear to encode the physical location of an agent within an environment[24], and this encoding appears to favor "important" areas of the environment, e.g., the platform in a water maze[25] or a junction in a t-maze[26]. Moreover, place cell representation of the environment is dynamic, responding to changes in structure or behavioral import[27,28]. Place cell function is critical for navigating an environment[29], and consistent with a role for DA, individuals with PD exhibit navigation deficits[30]. As with temporal representation, RL formulations generally represent locations within a physical space as discrete states that support learning, but not as the target of learning themselves (but see Stachenfeld et al., 2017[31]).

Finally, DA projections to PFC appear to play a critical role in working memory[1,5], categorization[32] and task representation[33]. The expression of value-based choice appears to rely on DA function[34], and tracking option values appears to be a central function of DA targets in lateral and medial PFC[35]. Computationally, models of basal ganglia and PFC have suggested that DA may underly working memory (WM) gating dynamics[36,37] governing whether and when an item should be maintained or expunged from WM. While the decision to store items in WM can be framed in terms of value optimization, the nature of *how* items are represented has received less attention. In models of PFC and WM, external stimuli frequently correspond to internal representations in a one-to-one fashion[36,38]: associative learning depends on the existence of these representations, but it is less clear how such representations emerge in the first place.

Thus, a common theme unifying the contexts in which DA is implicated, but which have not been satisfactorily addressed by the RPE hypothesis, is that of *representation*. In order for associative learning to occur, a representational substrate is required, and its form is frequently simplified to focus on the problem of learning associations: the representational space is generally static and defined to emphasize features that are presumed to be important for learning. Although these[39–42] simplifications may be relatively harmless when considering typical laboratory experiments, they may not apply to more ecologically valid situations in which a suitable representational space cannot be determined *a priori*. Indeed, it may be the case that sophisticated behavior *depends* on a malleable representational space.

If, as seems likely, internal representations used by animals and humans are updated during learning, an outstanding question concerns the mechanisms by which such representations are adjusted. Earlier work investigating attention and representation learning in RL frameworks proposes complementary signals to the standard RPE, or suggests how DA signals may support learning when to gate (but not necessarily learn) appropriate representations[37,38]. More recent advances in RL and machine learning demonstrate that sophisticated behaviors can be realized in systems in which the representational space is learned alongside the value function[43]. However, it is unclear to what extent representation learning in these systems corresponds with 1) how representations are learned in the brain via dopaminergic signaling, and 2) specifically how representations of state (as opposed to stimulus representations) are learned.

In applications of RL to training multilayer networks[44–46], including, for example, Deep Q-learning Networks[47] (DQN), the involvement of DA-like RPEs in representation learning is indirect: although RPEs serve as the underlying signal, representation learning frequently depends on the backpropagation of the RPE through multiple hidden layers to adjust connection weights. While this approach provides an end-to-end account of how RPEs can support complex task learning – an account which has proven extremely powerful in machine learning applications – it is questionable to what extent DA itself plays a role in learning representations from the earliest levels of stimulus representation, through intermediate state representations, to high-level response generation. This is especially problematic when considering that changes in the state representation of a task made in response to changing task contingencies can occur rapidly in the brain[48], at odds with the prolonged training needed to update representations in DQNs through incrementally adjusting weights.

Generally, state representations are correlated with – but nonetheless distinct from – stimulus representations. In the simplest cases, frequently modeled as a Markov Decision Process, stimuli and states have a one-to-one correspondence – each state is unambiguously specified by a stimulus. Under more realistic assumptions, modeled by partially observable Markov decision processes, stimuli are probabilistically associated with stimuli, and RL models within this framework attempt to infer the current state based on unreliable input. Although performing a task depends on the existence of suitable state representations, it remains unclear how such representations develop during learning and behavior.

Considering the well-documented role of DA in learning, the myriad contexts in which DA function is implicated in adjusting internal representations, and the speed with which DA influence brain function and behavior, we hypothesize that a major role for predictions errors reported by the DA system involves adjusting state representations in the brain. In this manuscript, we derive RL learning rules for adjusting internal state representations, and show how these rules can capture effects in which DA is implicated but are not directly addressed by associative learning.

## 2 | APPROACH

In applications of RL to modeling temporal, spatial and abstract cognitive processes, the distribution of states is typically defined by the modeler. For example, a transition between two states in a Markov chain can indicate the passage of a set amount of time, and transitioning between two adjacent points in a grid representation of the environment can reflect translation over a specific distance.

Although the assumptions of fixed representations distributed uniformly over some state space simplifies the modeling problem, there are good reasons to suspect that biological RL does not obey them. Evidence from single-unit recordings suggests that the temporal receptive fields of neurons are compressed immediately following the onset of a reward-predicting stimulus, while fewer neurons represent time periods far removed from stimulus onset[49]. Similarly, hippocampal place cells preferentially cluster around 'interesting' areas of the environment at the expense of other regions[25]. This tradeoff seems to make adaptive sense – a richer representation of reward-adjacent states may support more rapid and sensitive learning for critical regions of the state space, while a sparse representation of outlying states conserves resources without severely impacting performance.

Based on the observation that real state representations may be both malleable and adaptive, we propose that RPEs derived from RL may be useful not only for learning associations between state and reward, but also for learning optimal state representations.

The RL learning rule usually takes the form:

$$\Delta \mathbf{W} = \alpha \delta X \tag{1}$$

capturing the intuition that the mapping $\mathbf{W}$ between inputs with activity X and predictions is adjusted to reduce prediction errors ($\delta$): weights should increase for positive errors and decrease for negative errors, and this change is proportional to the magnitude of the error, reflected by a learning rate $\alpha$ (Fig. 1A) . Formally, eq. 1 is equivalent to taking the derivative of the squared prediction error with respect to the network weights.

Although changes in weights are the most common means of reducing error, alternative or complementary approaches may reduce error through adjusting the level of activity of a unit. Activity in units can be increased to reduce positive errors (assuming a positive associative weight) or decreased to reduce negative errors. If we assume a squared error as above, changes in unit activity can be determined using the derivative of the squared prediction error with respect to the function *f(input)*, mapping input to representation unit activity:

$$\Delta X = \alpha \times \delta \mathbf{W} \times f' input \tag{2}$$

Here, $\delta \mathbf{W}$ reflects the prediction error $\delta$ backpropagated through weights $\mathbf{W}$. When unit activity is a linear function of the input, this simplifies to $\Delta X = \alpha \times \delta \mathbf{W} \times constant$. Alternatively, the mapping function may be more complex; for the rest of the paper, we use a Gaussian function to determine unit activity:

$$X = e^{\frac{-input - \mu^2}{2\sigma^2}} \tag{3}$$

The Gaussian function is a popular way to model unit receptive fields: a unit is maximally active when the input occurs at a given value $\mu$, and activity decreases as a function of $\sigma$ as the input diverges from $\mu$. Changing representation unit activity to reduce error therefore depends on adjusting the parameters ($\mu$ and $\sigma$) of that unit's Gaussian, determined by taking the derivative of the squared prediction error with respect to each parameter. The partial derivatives of a Gaussian with respect to its underlying parameters are

$$\frac{\partial X}{\partial \mu} = e^{\frac{-input - \mu^2}{2\sigma^2}} \times \frac{-2}{2\sigma^2} \times -1 \tag{4}$$

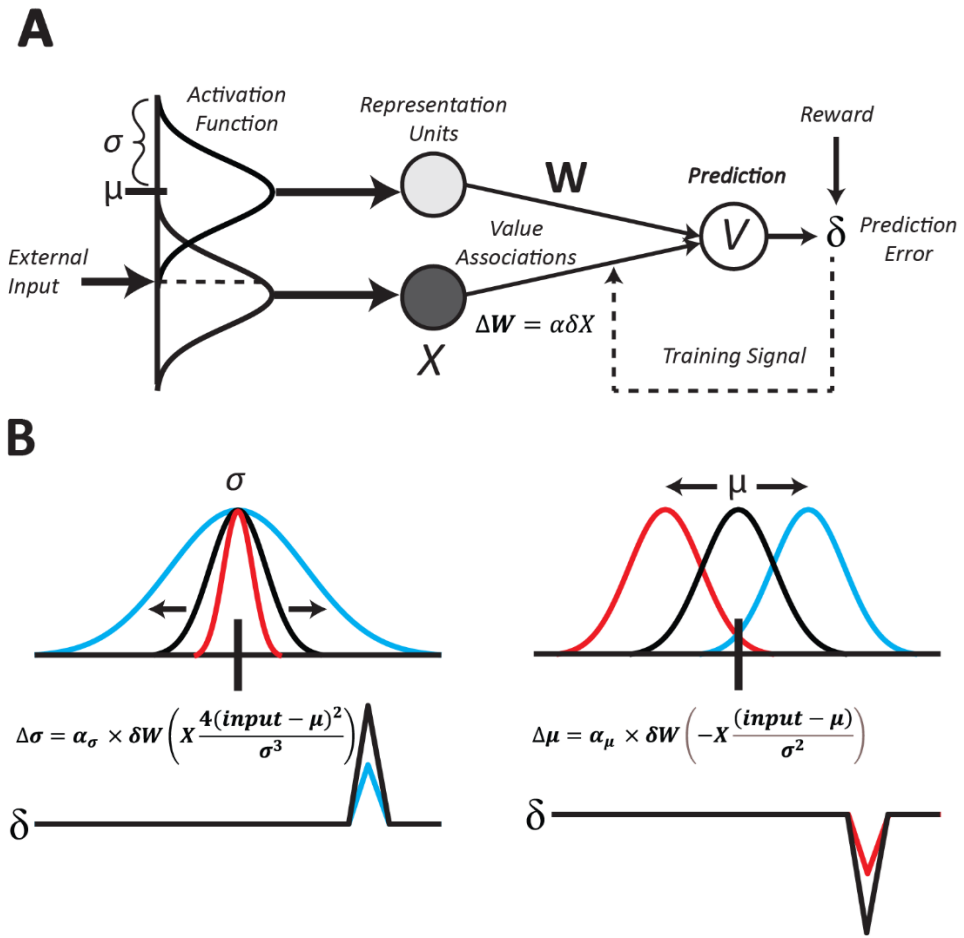

**FIGURE 1** Figure 1. General approach. A) Reinforcement learning models generally use reward prediction errors to train associations between stimulus representations and rewards. Internal stimulus representations are modeled as some function of external input – in this case, the strength of internal representations is determined by the representation unit's Gaussian activation function. **B)** Minimizing prediction errors can be accomplished by changes in the parameters of a unit's activation function as well. A positive prediction error (lower left frame) can be reduced by increasing the spread of the Gaussian (upper left, blue line) or moving the position of the Gaussian toward the prediction error (upper right, blue line). Alternately, negative prediction errors (lower right, red line) can be reduced by decreasing the Gaussian's spread (upper left, red line) or moving the Gaussian position away from the prediction error (upper right, red line).

$$\frac{\partial X}{\partial \sigma} = e^{\frac{-input - \mu^2}{2\sigma^2}} \times -input - \mu^2 \times \frac{-4}{\sigma^3} \tag{5}$$

Eqs 4 and 5 simplify to:

$$\frac{\partial X}{\partial \mu} = -X \frac{input - \mu}{\sigma^2} \tag{6}$$

$$\frac{\partial X}{\partial \sigma} = X \frac{4input - \mu^2}{\sigma^3} \tag{7}$$

Thus, in order reduce prediction error by changing the parameters of a Gaussian function, we get:

$$\Delta \mu = \alpha_\mu \times \delta \mathbf{W} \left( -X \frac{input - \mu}{\sigma^2} \right) \tag{8}$$

$$\Delta \sigma = \alpha_\sigma \times \delta \mathbf{W} \left( X \frac{4input - \mu^2}{\sigma^3} \right) \tag{9}$$

$\alpha_\mu$ and $\alpha_\sigma$ are learning rate parameters for mean and variance. In eqs 8 & 9, changes to $\mu$ and $\sigma$ are partially determined by the weight between an input and prediction. The use of weight information (sign and magnitude) is standard in artificial neural networks (e.g., error backpropagation), but the biological plausibility of learning in this manner remains in question. Insofar as our aim is to demonstrate how DA signals can support representation learning in the brain, our results should not depend on a mechanism for which there is not sufficient empirical evidence. Ignoring weight information in eqs 8 & 9 gives us biologically-friendly learning rules for adjusting Gaussian parameters.

$$\Delta \mu = \alpha_\mu \times \delta \left( -X \frac{input - \mu}{\sigma^2} \right) \tag{10}$$

$$\Delta \sigma = \alpha_\sigma \times \delta \left( X \frac{4input - \mu^2}{\sigma^3} \right) \tag{11}$$

Eqs 10 & 11 describe how the function mapping an input to unit activity can be adjusted by RPEs derived from RL formulations in order to minimize error. Although ignoring the weights in the learning rule appears to be rather drastic, we will show that the modified learning rule nonetheless still learns reasonable representations.

## LEARNING TEMPORAL REPRESENTATIONS.

To demonstrate how RPEs can be used to adjust the distribution and precision of temporal states, we adopt the standard temporal difference (TD) learning rule:

$$\delta_t = R_t + \gamma V_{t+1} - V_t \tag{12}$$

where R is the total level of reward at a given time $t$ and V is the prediction of future rewards discounted by $\gamma$ ($0 < \gamma < 1$) . We model time as a Markov chain composed of a series of sequential states where each state reflects the amount of time elapsed since the onset of some salient stimulus. In many applications, the current state of the process is known with absolute certainty, states transition with probability 1, and only one state is occupied at any instant, yielding tap-delay chain dynamics. A straightforward generalization of tap-delay chains is to introduce a Gaussian temporal receptive field[50,51] – the degree of activity of a representation is given by the both the time interval – coded by the mean $\mu$ – and variance $\sigma$. The activity $x$ of a temporal representation unit (TRU) at time $t$ is calculated as:

$$x_t = e^{\frac{-\left(t - \mu_x\right)^2}{2\sigma_x^2}} \tag{13}$$

Value predictions with Gaussian activity functions are computed as :

$$V_{t+1} = \sum_x x_t \times w_{x,t} \tag{14}$$

And equations 10 & 11 are adapted for time:

$$\Delta\mu_{x,t} = \alpha_\mu \times \delta_t \left(-x_t \frac{t - \mu_{x,t}}{\sigma_{x,t}^2}\right) \tag{15}$$

$$\Delta\sigma_{x,t} = \alpha_\sigma \times \delta_t \left(x_t \frac{4t - \mu_{x,t}^2}{\sigma_{x,t}^3}\right) \tag{16}$$

Although the rules apply locally to individual TRUs, our first series of simulations demonstrates how the interaction of TRUs under these equations determines how multiple TRUs are organized in time by RPE signals in order to minimize prediction errors. As with standard RL models, the model learns the discounted value of future rewards (Fig. 2A) by adjusting associative weights (eq. 1) to minimize RPEs (Fig. 2B). Additionally, the temporal receptive fields of TRUs in the model are updated by RPEs according to eqs 15 & 16, resulting in units that are distributed throughout the interval between a reward-predicting stimulus and the reward it predicts, as well as unit activity profiles with sharper or broader peaks (Fig. 2C).

This distribution of TRUs acquires two characteristics during reward learning (Fig. 2D): first, after learning, time periods around salient events, such as the onset of a predicted reward or a reward-predicting stimulus, are *more densely* represented by TRUs than are intervals with no salient events. Second, the activity profile of TRUs immediately following stimulus presentation is *sharper* than the profile of units occurring later in the interval. This reproduces patterns of activity observed in neurons in putamen corresponding to the temporal representation of stimuli (Fig 2C)[49].

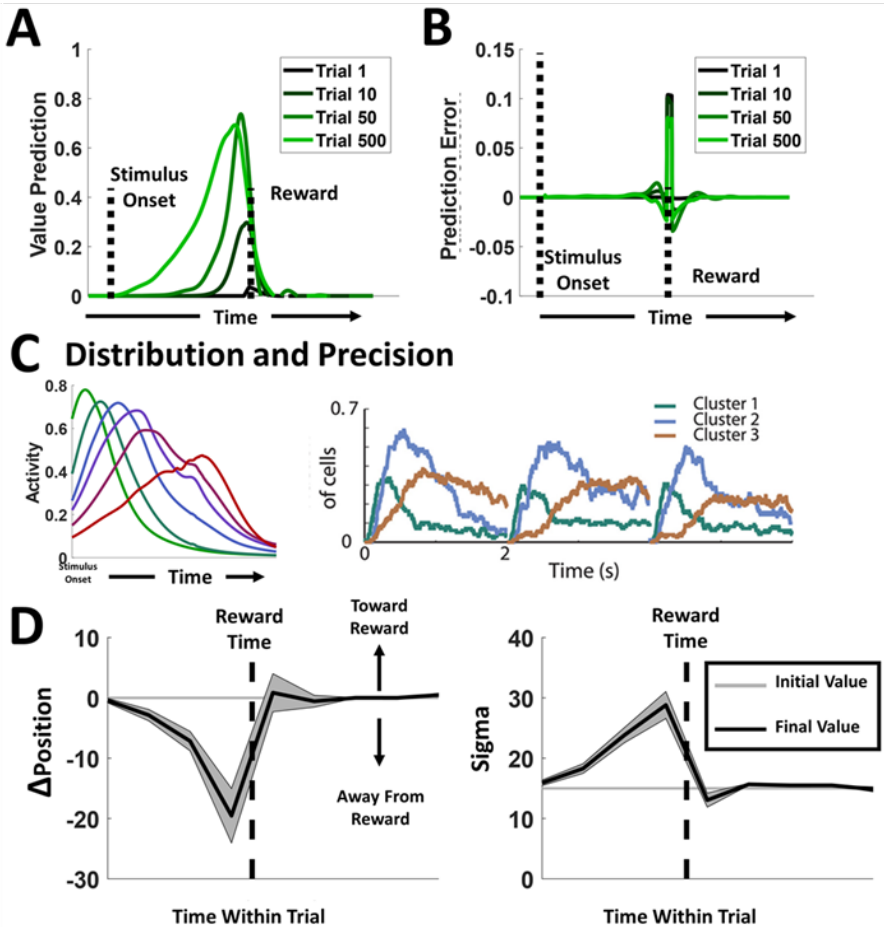

**FIGURE 2** Figure 2. Temporal Representation. A model in which the position (temporal interval) and variance of representation unit are malleable captures key characteristics of midbrain timing units. **A)** As in standard temporal difference learning, the model learns a discounted value prediction from the onset of a stimulus peak at the time a reward is expected. **B)** As reward predictions converge, the prediction error at the time of reward decreases over repeated trials. Because the activity of temporal representation units is modeled with a Gaussian activation function, and the Gaussians for neighboring units may overlap, value predictions are noisy and may in some cases exceed the total possible reward value. This overlap can result in transient increases in reward prediction errors during learning. **C)** In order to reduce prediction errors, the distribution and variance of temporal representation units is adjusted over time. Following training **(left)**, units active immediately following the onset of a reward-predicting stimuli have a more temporally precise activation pattern than subsequently active units. This pattern matches well with data from single units in putamen showing apparent spectral timing patterns of activation (**right,** reprinted from ref.[49] with permission). **D)** Starting from a uniform distribution (**rightr, gray line**) over a time interval covering both the stimulus and reward events, and a uniform gaussian variance (**right, gray line**), temporal units are redistributed away from the time at which a reward occurs (**left, black line**), and their temporal precision varies with the interval length (**right, black line**). Gray margins represent the standard error of the mean.

If we take the mean firing interval of a TRU to represent the 'tick' of an internal clock, these results are consistent with DA's influence on temporal perception . Prior to learning, each tick of the clock marks off a fixed, randomly determined temporal interval, while after learning, clock ticks are pulled toward the stimulus onset – there are more ticks of the internal clock in a given amount of absolute time following a reward-predicting stimulus. DA neurons in SNc, absent exogenous demands, fire in a steady "pace-maker" fashion both *in vitro* and *in vivo*[52]. Changes in the temporal profile of DA activity due to reward or reward-predicting stimuli away from pace-maker dynamics may provide a signal for adjusting the position and receptive field of TRUs. This manuscript proposes that the DA signal is useful for coordinating neural activity over time, and that DA depletion, either pharmacologically or through disease, may interfere with such coordination by hampering effective DA entrainment of TRUs.

A role for DA in adjusting the profile of temporal receptive fields as well as value associations has previously been proposed[53]. In contrast to the present study, however, receptive field properties were adjusted indirectly through setting the frequency of an internal clock rather than learned directly. 'Centralized' control of a global clock frequency may provide a more intuitive account for phenomena that seem to require near-instantaneous adjustment of temporal receptive field properties, e.g., dopamine-related time dilation following blinks[54]. Although both approaches reproduce firing patterns of timing neurons in putamen[49], the current approach may generalize more readily to non-temporal contexts such as spatial navigation and classification problems, as discussed below. One possibility is that DA signals may influence global parameters, such as internal clock frequency, while also supporting local adjustments in receptive field properties.

## MOTOR CONTROL

Entrainment of TRUs is one mechanism through which DA may influence motor behavior. Bursting DA activity is observed prior[55], and is causally related[56], to movement initiation, and motor-related bursting activity may play a similar role as RPEs in supporting learning[57]. Salient external stimuli are known to elicit DA responses[58], and the use of regular repetitive stimuli (e.g., metronomes) has been found to improve motor symptoms of PD – such as the irregular gait of advanced parkinsonism – that rely on rhythmic coordination[59]. The implication, then, is that the same neurons reporting PEs related to salient stimuli may also underlie initiation of a motor behavior, and that bursting DA activity prior to the generation of a motor output may be used for maintaining learned temporal distributions underlying rhythmic behavior.

Together with our previous results (Fig 2), these observations suggest that 1) DA bursts related to a salient stimulus may play a similar role in adjusting the timing and distribution of TRUs, especially when the salient stimulus is predictable (e.g., presented repeatedly at constant intervals), 2) DA neurons that respond to salient stimuli also drive motor behavior, 3) DA bursts derived from motor responses may play the same functional role as DA bursts derived from salient stimuli in learning and maintaining temporal representations. Point 3 further suggests that 4) the symptoms of diseases that target the DA system, such as PD, result from the inability to develop and maintain stable temporal representations.

In motor timing studies including individuals with PD, PD subjects were able to synchronize motor behavior (finger tapping) with a salient external cue (metronomes) but exhibited deficits when required to continue the behavior absent the external cue. Specifically, inter-response intervals are frequently significantly more rapid for the PD group during the continuation phase of a synchronization-continuation task than for control groups (reviewed in ref.[60]). To test whether the organization of TRUs might account for this effect, we simulated the model on a synchronization-continuation experiment. For each tick of a simulated metronome, a DA impulse was generated. In these simulations, the DA impulse served both as a US and CS (see simulation methods). As in our simulations of simple stimulus-dependent reward prediction above, during the synchronization period TRUs clustered around the onset of the impulse (Fig. 3, top left) and more precisely represented those periods relative to periods in between impulses.

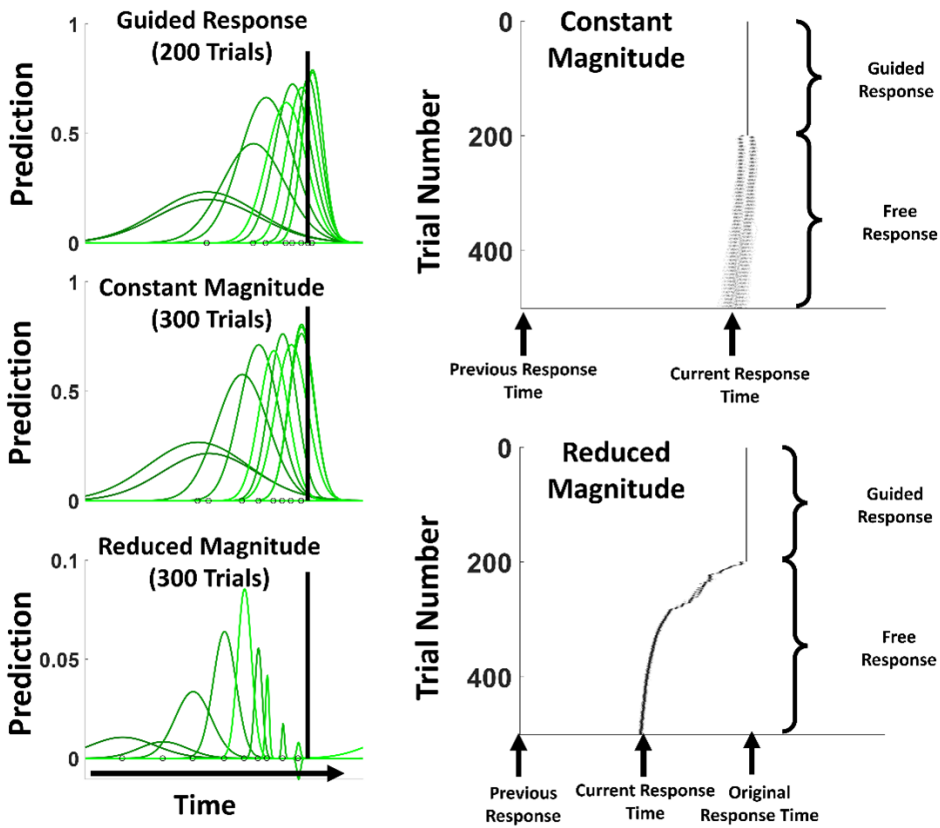

**FIGURE 3**    Figure 3. Timing and motor response. The model is trained to predict a reward delivered at regular intervals for 200 trials (upper left frame), and each reward serves as the CS for predicting subsequent rewards. After this initial learning period, the model's own predictions serve as the basis for future rewards: a rewarded 'response' is generated at the peak of the value prediction. When the reward generated by model's response is the same magnitude as the reward during the trained period (center left, upper right), model predictions remain relatively stable over time and the frequency of model responses drifts only slightly. In contrast, when the self-generated reward magnitude is significantly lower (lower left, lower right), the frequency of responses increases rapidly.

Following the initial synchronization period, the exogenously generated DA impulse was discontinued. Instead, the model was required to generate its own DA response through a motor output (under the assumption that motor behavior-derived DA bursts serve the same functional role as bursting DA activity driven by other sources). When the model's self-generated DA response remains high, it is capable of maintaining the trained frequency over hundreds of trials (Fig. 3, center left, top right). However, when the self-generated DA response is low, temporal precision decreases and the frequency of model-generated motor output increases, mirroring the increased cadence observed in empirical data (Fig. 3, bottom left & bottom right). These changes in temporal representation and motor behavior derive from the weaker-than-expected DA signal resulting in a negative PE at the usual time. In order to reduce the negative PE, the centers of TRU receptive fields are pushed away from (earlier than) the time at which the negative PE occurs (eq. 15), resulting in a shorter interval between successive motor responses.

Although these simulations are suggestive of how DA's role in temporal representation may translate to motor deficits, we are not aware of any experiments that have directly tested this hypothesis. Here our simulation results depend on an attenuated RPE in the "Reduced Magnitude" condition (Fig. 3, lower left) that increases response frequency. This suggests a possible approach for testing our hypothesis against the "global clock" frequency[53] hypothesis mentioned above. Specifically, under the global clock hypothesis of DA's influence on temporal representation, attenuated DA levels produce longer temporal intervals, which would translate to a lower overall frequency in a synchronization-continuation task. Direct manipulation of DA activity in an animal model of the synchronization-continuation task[61] may be able to provide evidence in favor of one or the other accounts.

## SPATIAL REPRESENTATION

Besides being able to account for effects related to the representation and estimation of time, the approach outlined here translates easily to spatial representation contexts. Whereas a role for dopamine in, for example, updating the speed of an internal clock could apply to spatial contexts by incorporating additional assumptions (e.g., interpreting spatial distances in terms of the time needed to traverse them), using RPEs to update the location and variance of spatial representations barely requires an update of our notation. Instead of identifying states by their mean time interval and variance, we can instead define states by their two-dimensional coordinates and variance. Learning rules for updating the 2D location of spatial representation units (SRUs) are essentially identical to the temporal case. Here $i$ and $j$ refer to vertical and horizontal dimensions:

$$\Delta\mu_{x,i} = \alpha_\mu \ \times \ \delta \left( -x \frac{position_i - \mu_{x,i}}{\sigma_{x,i}^2} \right) \tag{17}$$

$$\Delta\mu_{x,j} = \alpha_\mu \ \times \ \delta \left( -x \frac{position_j - \mu_{x,j}}{\sigma_{x,j}^2} \right) \tag{18}$$

$$\Delta\sigma_{x,i} = \alpha_\sigma \ \times \ \delta \left( x \frac{4 position_i - \mu_{x,i}^2}{\sigma_{x,i}^3} \right) \tag{19}$$

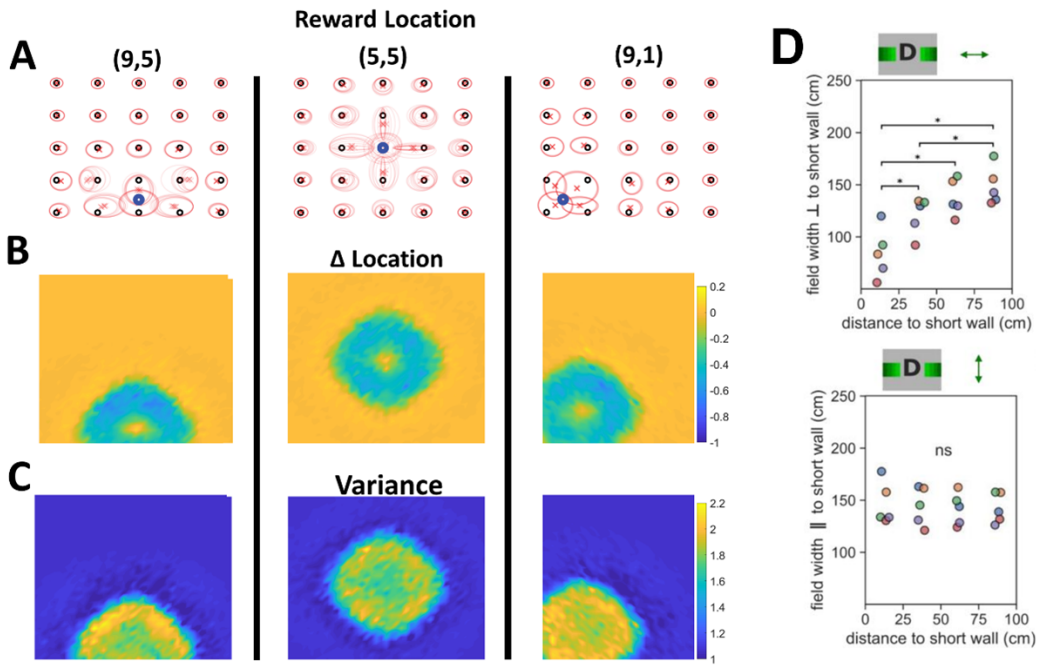

**FIGURE 4** Spatial Representation. We simulated a reinforcement learning agent in a simple grid environment with a single reward source at the edge, center, or corner (left, center, right columns). **A)** Simulations in which representation units were distributed uniformly through the environment (black circles) prior to learning demonstrate the adjustment of position (red Xs) and spread (red circles) for 2D gaussian activation functions. Units move toward the reward source, while the spread of the activation function increases along the axis representing the most valuable path and decreases along the orthogonal axis. **B)** Initiating unit positions to random locations illustrates the region around the reward source in which units are attracted (cool colors) to the reward location. **C)** Representation units immediately proximal to the reward location also tend to be more precise than those less close, while the precision of units far from the reward source does not change from the initialized value. **D)** The distribution of place cells in rat hippocampus reflects visual information in the environment – place cells are more densely clustered around regions with higher visual change and have narrower receptive fields relative to regions of the environment with lower visual change. This pattern is reproduced in our simulations but using change in reward rather than visual information. Figures originally published under a CC BY license allowing reuse with adaptation.

$$\Delta\sigma_{x,j} = \alpha_\sigma \ \times \ \delta \left( x \frac{4position_j - \mu_{x,j}{}^2}{\sigma_{x,j}^3} \right) \tag{20}$$

Simulations using an actor-critic architecture in a simple grid environment demonstrate how SRUs cluster around areas of the environment associated with reward (Fig. 4A), similar to the over-representation of reinforcing areas of the environment in hippocampal place cells[25]. In contrast, areas of the environment without behaviorally relevant import are more sparsely represented. SRUs initially located close to reward sources in the environment (Fig. 4B) are pulled toward the reward source during training, while the position of SRUs located far from the source remain mostly unchanged. Regions around the reward source are represented more precisely than regions farther away (Fig. 4C). Values of $\sigma$ for SRUs immediately adjacent to a reward source are lower than for SRUs slightly farther away, while SRUs positioned very far from the reward source, $\sigma$ remains largely unchanged from their initial values.

In this respect, empirical evidence disagrees with the our simulations; although hippocampal place cells are observed to cluster around goal or reward states in the environment, the size of their receptive fields is not observed to change[25], neither becoming broader[31] or narrower. In translating the learning equations derived for temporal representations to a spatial context, we included learning rules for adjusting both location and size. However, location and size updates are independent processes in the model; learning rules for updating field size (eqs 19 and 20) could be removed without affecting rules for updating location. This aspect notwithstanding, these simulations demonstrate that a signal generally considered to support adaptive behavior through associative learning can also be leveraged to support the adaptive allocation of limited representational resources.

Although the size of place cell fields has not been observed to be sensitive to reward, our results provide an interesting point of contact with recent findings examining the influence of visual information on place cell precision and distribution within an environment[62]. Specifically, it was observed that place cells in regions of an environment in which visual change is low (e.g., near the center of a walled environment where movement does not produce large differences in the visual scene) were more sparsely distributed with larger fields. Conversely, place cells were more densely distributed, with smaller place fields, near regions in which movement resulted in large visual changes, such as movement toward a boundary wall. Our simulations find much the same effect, with the exception that instead of visual change, SRUs in our simulations distribute themselves according to changes in predicted reward: when far from a rewarding location, movement towards the reward produces only small increases in the reward prediction (eq. 12), while the same movement produces a larger increase when the rewarding location is already close by.

## CATEGORY REPRESENTATION

The immediate generalizability of our approach from a temporal to a spatial context further suggests that it may be useful for learning optimal internal representations for abstract categorization problems. We adapted the actor-critic model used for simulating spatial representation to perform a 1- (Fig 5A) and 2-dimensional (Fig 5B) classification task. Input to the model consists of continuous numbers. In the 1D case, input values below a decision bound are assigned to one category while values above are assigned to another, and the model must learn to respond with the appropriate category. Because each trial consists of a single input and a single response, the TD error simplifies to:

$$\delta_t = R_t - V_t \tag{21}$$

i.e., there is no prediction of future rewards. All other learning rules are identical to those used in our spatial representation simulations.

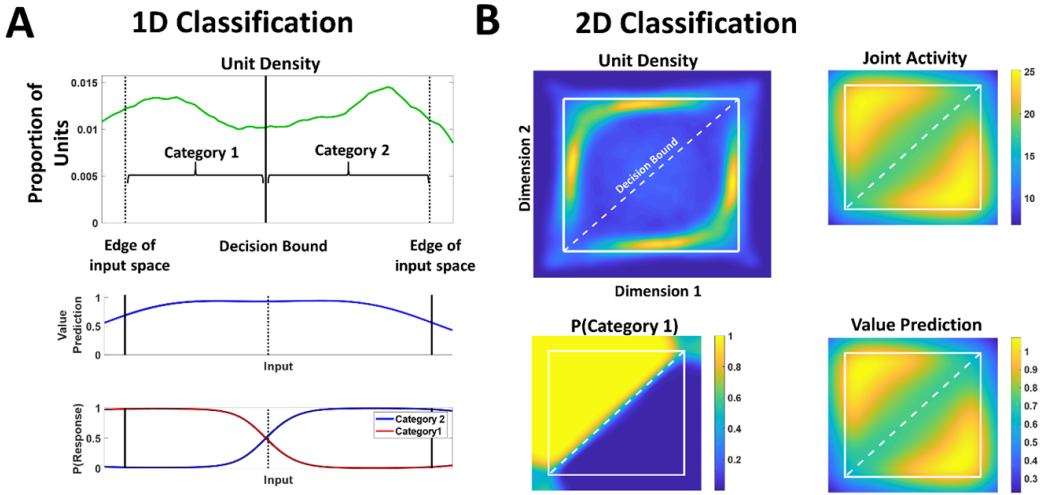

**FIGURE 5**   Category Representation. Using prediction errors to adjust spatial position of units in a categorization task shifts global attention. **A)** In the case of a one-dimensional categorization task, regions close to the edges of the input space (top row) are over-represented compared to regions near the decision bound. Despite this over-representation, value predictions (middle row) and behavior (bottom row) are relatively uniform for each category. **B)** These effects hold in the 2D case. Here, distinct clusters of units emerge along each of the edges defining the input space (upper left frame), while summed unit activity (upper right), attendant value predictions (lower right), and behavior (lower left) are more evenly distributed throughout the category areas.

     Representation units were initialized to random locations drawn from a range extending slightly beyond the space of possible inputs. During training, units tended to redistribute themselves away from the decision bound and toward the regions of the input space near the outer edges of the two categories (Fig. 5A). Movement away from the decision bound is a product of eqs 17 & 18: samples close to the decision bound are more likely to produce incorrect responses and negative predictions errors – in order to minimize the negative prediction error, the unit relocates farther away from the bound, and the width of its receptive field shrinks.

     This effect is also observed in the two-dimensional case (Fig 5B), where the decision bound is a line and determining the correct category depends on integrating the values of two features. As in the one-dimensional case, following training the edges of the input space are over-represented while regions near the bound are under-represented. Although individual units tend be maximally active for samples drawn from the edges of the input space, the *global* response (i.e., the summed activity of all units) tends to be maximum for samples drawn closer to the center of a category's distribution. That is, although only a relative few units respond strongly to the 'prototype' of a category, *many* units respond weakly.

     While the learning rules for updating receptive field location and size yield representations consistent with positive evidence for a category, other lines of work suggest that individuals attempt to identify the decision bound separating categories. Models derived from Decision Bound Theory[63] have been used to account for distributed brain activity during category learning[64], and individuals adopt information-sampling strategies that probe decision boundaries[65]. Recent work has identified neurons in monkey pre-SMA[66] and conflict-sensitive neurons in supplementary eye field[67] with activity consistent with decision bound representation.

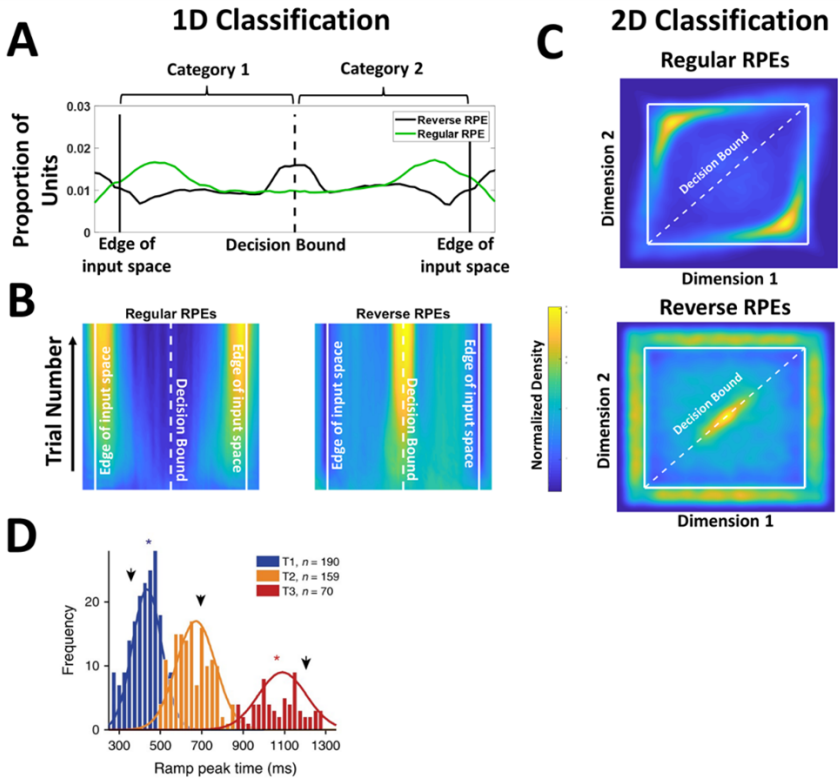

**FIGURE 6** **A)** Representation units trained by 'reversed polarity' prediction errors cluster around the decision bound. **B)** Movement of 'normal' and 'reversed' representation units toward the edges and center of the input space, respectively, develops simultaneously during training. **C)** These effects are again reflected in the 2D case, with a higher density of 'regular' representation units along the internal edges of the input space, and 'reversed' units distributed along side the decision bound. **D)** Example boundary neurons recorded from monkey pre-SMA during a temporal discrimination task (reprinted from ref. [66], published under a Creative Commons Attribution 4.0 International License, which permits adaptation with attribution.)

In our framework, traditional RPEs promote the distribution of representation units away from the decision bound, i.e., away from locations where negative prediction errors are more frequent. Although RPEs signaled by midbrain DA neurons are generally valenced (positive RPE = positive valence), it has become increasingly clear that some subpopulations code RPEs in which the polarity of the valence is reversed (positive RPE = negative valence)[58]. If the same learning rules for updating receptive field position and variance apply to 'reversed-polarity' RPEs, would this explain effects related to decision boundary representation?

In order to test this, we repeated our category learning simulations as above. However, for half of the representation units (randomly selected) the sign on the error signal was flipped for eqs 17-20 (the sign of the RPE for learning associative weights remained the same, however). As before, units that updated their position and variance using the traditional RPE were distributed toward the edges of the input space. In contrast, units with polarity-reversed learning were over-represented both outside the edges of the input space (pushed there by positive PEs generated for easy samples from one category or the other) as well as along the decision boundary that optimally separates categories.

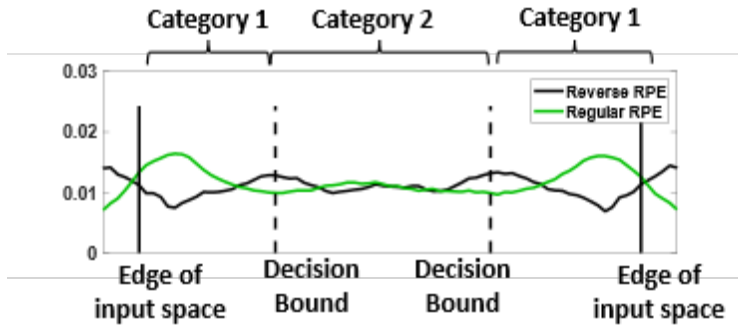

**FIGURE 7** **Linear inseparability.** Simulations of the model on a linearly inseparable classification problem recapitulate results from linearly separable problems (Figs 4 &5). Units in the model trained with normal RPEs tend to cluster along edges of the input space, while units trained with reversed RPEs cluster at the decision boundaries.

The above simulations deal with categorization problems that are linearly separable, i.e., those that require only a single hyperplane to divide the space between examples of one class from the other. In order to test whether our approach also extends to linearly inseparable problems, we conducted an additional simulation for the 1D case in which samples from the input space between 3 & 6 belonged to one class, and those below 3 or above 6 to the other. As in our simulations for linearly separable problems, the model learns to identify edges of the input space as well as decision boundaries (Fig. 6). The ability of the model to solve linearly inseparable problems depends more on the existence of multiple state representation units than on the dynamics by which the receptive fields of those units are adjusted. Assuming the state space is adequately covered by static state representation units, adjusting their position and width is not necessary. However, updating receptive fields allows state representation units to identify interesting areas of the state space that might otherwise not be immediately apparent based on units with fixed positions.

## 3 | DISCUSSION

In this manuscript, we describe an approach for using scalar RPEs derived from RL formulations to adjust the parameters of a unit's activation function. In doing so, we provide solutions to several outstanding issues concerning the role of DA in learning and adaptive behavior. By using RPE signals to adjust the location and spread of internal representations of external stimuli, we are able to account for effects related to temporal and spatial representation, motor control, and abstract decision making. While these effects have all been observed to be influenced by DA (dys)function, standard RL formulations focusing exclusively on value learning have not been able to easily accommodate them.

By extending the role of RPEs to adjusting parameters of an activation function, our approach harkens back to early associative[68,69] and connectionist[70] models of attention. Generally, attention can be described as the prioritization of some subset of stimuli or stimulus features, usually at the expense of others[71]. Given this definition, attention-like processes are observed at two levels under our approach. First, the learning rules for updating the center and variance of receptive fields of units in the model correspond to low-level attentional processes. At the level of single neurons, DA is implicated in adjusting the signal-to-noise (SNR) ratio of target neurons[72]: a neuron with high SNR is active only when stimuli provide a good match for its preferred input, and silent otherwise, while a low SNR neuron responds weakly to a broad range of inputs. During model training, the variance of the receptive fields for individual units shifts in response to RPE sign – expanding for positive PEs and contracting for negative PEs. Simultaneously, the center of a unit's receptive field moves toward positive RPEs and away from negative RPEs (eqs 10 and 11). Thus, attention is realized locally by adjusting the receptive fields of individual units to minimize prediction errors.

While individual units attempt to minimize local PEs, global attentional effects emerge from indirect interactions amongst representational units. When a positive PE occurs in the overlap of two or more units' receptive fields, all units attempt to minimize that error through update of associative weights (eq 1) and adjustment of the center and variance of their receptive fields (eqs 8 & 9). Frequently it will occur that, on subsequent trials, the combined value predictions of those units will then exceed the total reward that generated the PE (e.g., Fig. 2A) to begin with, resulting in a negative PE and leading to weaker associative weights and further adjustments in the center and variance of the units' receptive fields. The final distribution of all units is thus a product of local learning rules for minimizing positive and negative PEs. At the global scale, this produces a characteristic distribution of units that favors denser and more precise representations around regions of the input space that produce positive RPEs, i.e., the model learns to preferentially represent 'interesting' areas of the space.

The adjustment of the size and location of receptive fields for states mirrors the role of top-down attention in sensory processing, where the location of peak firing and size of early sensory receptive fields is malleable[73]. Although, at least within midbrain structures, the role of DA seems to be direct – DA neurons in SNc directly project to putamen[74], for example, with immediate effects on activity of target neurons – it is unlikely that DA PE signals themselves directly adjust sensory receptive fields. Regions that are implicated in top-down adjustment of sensory receptive fields[75] are themselves innervated by the DA system, and neurons that project to sensory areas from these regions express DA receptors[76]. One possibility is that DA PEs might provide a signal that indicates under what circumstances regions that more directly update low-level tuning parameters do so.

An open question in computational neuroscience and RL concerns learning not only the value function for a given set of state representations, but also learning state representations for a given task[77]. The density and distribution of state representations reflect a tradeoff between adequately characterizing the value function for a task vs. conserving resources. An exhaustive state representation may allow an exact characterization of the value function, but comes with heavy computational overhead, while a sparse representation is less computationally demanding, but also less precise. Our approach suggests how this tradeoff can be navigated using biologically plausible RPEs for learning both the value function as well as the state representation simultaneously. By adaptively distributing state representations to regions of the input space that are 'interesting' and away from less interesting regions, our approach provides a solution to the problem of optimally distributing limited attentional resources.

For our categorization simulations, the 'interesting' areas of the state space along which state representations distribute themselves during learning correspond to boundaries of the state space as well as decision bounds between categories. Besides mapping input to abstract categories, the same processes underlying the identification of decision boundaries may generalize to other contexts. Recent work in cognitive and developmental neuroscience examines how the continuous stream of sensory and perceptual input is experienced as a sequence of discrete events. Segmenting events appears to be critical for structuring memory and learning, but the ability to do so is itself something that must be learned – sensory input does not come 'tagged' with event information. In much the same way that the learning rules we derive for updating temporal receptive fields (eqs 8 & 9) translate to spatial contexts (eqs 17-20) with minimal changes, it may be the case that the learned distribution of representations that emerges in our categorization simulations (figs 4-6) also generalize to support learning distributions that identify boundaries between temporally extended events. One possibility is that events are defined by the similarity of information belonging to one event relative to information belonging to another. If DA plays a role in learning event boundaries, our model suggests that DA bursting activity should be observed when successive observations are highly dissimilar. Consistent with this idea is the finding that DA neurons appear to encode sensory prediction errors[78], but it remains to be seen whether this supports event segmentation.

An ongoing research concern is to explain the role of the DA signal across the range of functions in which it is implicated. The DA signal projects to a broad array of regions[79–81], including areas in prefrontal cortex, anterior cingulate, hippocampus, and anterior insula. These regions are implicated in representing different aspects of a decision task, including, e.g., state variables[38], response-outcome contingencies[82], or affective import[83]. Previous proposals have suggested that the DA signal may be vector-valued rather than the scalar signal commonly assumed in RL models[84,85]; the apparent functional specialization of projection targets could then be regarded as a product of receiving different components of a multidimensional error signal. While there is evidence that DA neurons in VTA and SNc are topologically organized by projection targets[86] and exhibit topologically organized functional diversity[87], it remains to be established that this topological organization corresponds to a multidimensional RPE. One possibility, suggested by our simulations, is that DA neurons coding traditional RPEs project to different targets than "polarity-reversed" RPEs, and thus support distributed representations of categories according to exemplar or decision bound theories.

The idea of adjusting receptive field location and variance has previously been used in models of human category learning, especially within connectionist frameworks. Our approach bears a resemblance to three such models, SUSTAIN[88], ALCOVE[89], and EXIT[70]. Unlike SUSTAIN, wherein negative feedback results in the formation of new clusters in the state space, negative feedback in our model results in adjustments of existing receptive fields. That is, SUSTAIN deals with errors by increasing the number of representation units, while our approach maintains the same number but adjusts the distribution of the units across the state space. Additionally, in our approach, the rules for such adjustments apply to all units in the model at all times rather than to a single winning cluster as in SUSTAIN. Our formulation of adjusting receptive field size is closely related to that of ALCOVE in that it attempts to reduce error through gradient descent (albeit slightly modified for biological plausibility). Unlike our approach, however, ALCOVE was trained using supervised learning, and the position of exemplar nodes in the network did not change during learning. A version of ALCOVE, Q-ALCOVE[90], substitutes the supervised classification error with a TD error and to show how receptive fields in a gridworld task shift during learning. While the use of a TD error is similar to our own approach, the location of receptive fields in Q-ALCOVE remained static as in the original. A previous RL model implemented an attention gain process similar to that formalized in EXIT to explore how learning could be facilitated by identifying relevant features in a high-dimensional feature space, but did not extend to adjusting the internal representation of the space[91]. To the best of our knowledge, the approach outlined here is the first that combines continuously active RL rules to update the distribution of state representations.

## 3.1 | Limitations & Extensions

Despite the array of results that can be accommodated by our framework, there is as of yet little direct evidence supporting the possibility that DA RPEs directly adjust receptive field properties of neurons in projection targets. While there is evidence that catecholamines, including DA and noradrenaline, are implicated in altering the SNR of neuronal response[92], it is unclear whether this alteration is the product specifically of phasic activity associated with PEs or instead results from, e.g., variations in tonic neuromodulatory levels. Furthermore, although the long-term properties of neuronal receptive fields can shift in response to changes in task contingencies[93], it may be the case that the action of neuromodulators produces only transient adjustments rather than directly effecting stable and long-lasting changes. Additional work is needed to explore the implications of the framework presented in this manuscript.

The learning rules we derive for adjusting location and spread of receptive fields depend on explicit representation of the absolute position of a unit in an environment as well as the variance of the Gaussian function governing its response sensitivity. It seems unlikely that single units would have access to this kind of direct information regarding their own receptive field properties. However, it is plausible that the key intuitions underlying the learning rules could be realized with a more biologically plausible architecture.

Furthermore, in deriving our learning rules, we made simplifying assumptions about the mapping of input to unit activation and its generalization to multidimensional spaces. In our simulations, we adopted a Gaussian function to map model input to representation activity. Gaussian activation functions are frequently used to model input that has a ready spatial interpretation. Models of early visual cortex, for example, sometimes use Gaussians to describe the firing characteristics of individual neurons[94]. It is questionable whether the assumption of a Gaussian activation function applies to all contexts, especially to, e.g., abstract categorical spaces. In principle, however, the approach described in this manuscript could be applied to other activation functions as well.

Additionally, for the 2D case, we use a simplified Gaussian function that ignores possible relationships between dimensions. That is, changes to $\mu$ and $\sigma$ occur along the axes of the dimensions. More complex learning rules would also consider the covariance between dimensions, and may include adjusting the covariance matrix in addition to mean and variance. Doing so would provide a mechanism for updating not only the location and spread of a receptive field, but also its orientation.

Our simulations use extremely simplified 1- and 2D Gaussians defined on a continuous input space as the bases for state representations. It is likely that state representations in the brain can consist of many more dimensions and some of these dimensions may not be well-captured by Gaussian functions (e.g., discrete or categorical inputs). Although the learning rules we derive in this manuscript can be extended to arbitrarily many dimensions relatively easily, incorporating information about covariance amongst dimensions during learning, as we suggest above, may render the framework unwieldy. However, it remains to be seen whether and under what circumstance the assumption of a Gaussian function is valid.

Furthermore, it is not clear how the learning rules described here would work for the categorical input case, where mean and variance parameters are not defined. One possibility is that distances between samples from a categorical dimension could be defined using the output space as a metric: samples that predict similar consequences are more similar to one another than those that predict different consequences. Although only a preliminary suggestion in this context, the notions of *acquired distinctiveness* and *acquired equivalence* have a long history in psychological research[95–98]. It remains to be seen to what extent they may be applicable to learned state representations.

As our approach derives from an RL formulation, numerous repetitions are needed for learned values to converge. While it seems plausible that extensive experience with a given task could result in gradual, long-lasting shifts in the distribution and precision of receptive field properties, this may be at odds with evidence from, e.g., Parkinson's patients exhibiting almost instantaneous shifts in behavior in response to administration of L-DOPA[49]. One reason for this discrepancy may be that in our approach, value associations are learned in conjunction with activation function properties – in order for changes in the location or spread of a receptive field to be effective in minimizing prediction error, value associations need to be at least partially established for changes in receptive field properties to have any appreciable effect.

Finally, an important goal for future work is to distinguish our account from alternative theories of representation learning (including theories based on supervised and unsupervised learning). This will require fine-grained studies of representational dynamics during learning, manipulations of feedback, and perturbations of the dopamine system.

## 3.2 | Conclusion

In summary, our interpretation of DA RPEs as adjusting the location and variance of receptive fields of neurons in target projection sites complements the traditional role of DA in driving associative learning. In doing so, we are able to account for the role of DA in contexts that are not easily accommodated by traditional RL approaches. Of particular note is that our interpretation does not require extensive revision of the RPE hypothesis of DA function – the same RPE signal used to learn value associations can also be used for learning temporal, spatial, and category representations. The myriad roles in which DA is implicated can thus be viewed not as different functions of the DA system itself, but as deriving from a general system for signaling PEs applied to diverse inputs.

## ACKNOWLEDGMENTS

This material is based upon work supported by the Air Force Office of Scientific Research under award number FA9550-20-1-0413.

# 4 | SIMULATION METHODS

## 4.1 | Temporal Representation

Eleven temporal representation units were used in these simulations. Units were initialized with a sigma of 15 and mu values from 0 to 825 time steps (randomly distributed throughout the interval with uniform probability.) $\alpha_\mu$, $\alpha_\sigma$, and $\alpha_W$ were set to 0.2, 0.05, and 0.1, respectively. The temporal discount rate $\gamma$ was set to 0.99. A single simulated trial consisted of 850 iterations. At 50 iterations, a reward-predicting stimulus was presented and remained for the duration of the trial. At 350 iterations, a reward was presented. The total, non-discounted value of the reward was equal to 1, and the duration of the reward presentation was 10 iterations. Thus at each iteration of the reward presentation, the current reward level was 1/10. A learning run consisted of 500 simulated trials, during which the model learned predictions of future reward level as well as unit-specific sigma and mu values as detailed in the main text. 50 simulations were conducted and results in fig. 2 reflect averaged values over these runs.

## 4.2 | Motor Control

For motor control simulations, we used a two temporal delay chains of 10 units each spanning 500 model iterations. Accordingly, temporal representation units were initially distributed at intervals of 50 iterations (sigma values were initiated to 75). The model's task was to learn the interval between reinforcement signals ($R_t$ in eq. 12), conceived to reflect a salient external input such as the tick of a metronome. Each 'tick' served as the US for one of the delay chains and the CS for the other: every odd tick initiated the first delay chain, and every even tick the second.

200 initial trials were used to train the model using the 'ticks'. During this training period, a 'reward' of magnitude 5.5 and length 10 iterations (reward/iteration = 0.55) was presented at the 250[th] iteration following the previous reward(i.e., rewards occurred at an interval of 500 iterations) . Following the training period, learned value predictions in the model were used to generate responses (300 trials total). For control simulations, model responses resulted in a reward of similar magnitude and duration. For Reduced Magnitude simulations, the reward magnitude was set to 0.1, with the same duration. Model response times were determined by taking the average iteration value for iterations in which value predictions were above the 99[th] percentile.

This procedure was repeated 1000 times, and figures display the averaged values.

## 4.3 | Spatial Representation

We implemented an actor-critic RL agent to navigate a grid world in search of reward. The critic component received input from spatial representation units (SRUs) with activity determined by the current position of the agent in the environment and the gaussian activation mapping external position to internal representation. Associative weights in the critic were updated according to eq. 1. The actor component received input from the same SRUs, and used that information to determine one of four possible actions (A), corresponding to translation in a cardinal direction (up/down/right/left):

$$A = X \bullet W_A$$

where $W_A$ are weights mapping input X to A. Actions were selected probabilistically using a softmax function.

$$P\left(A_i\right) = \frac{exp\beta A_i}{\sum_j exp\beta A_j}$$

Actor weights were trained using the prediction error generated by the critic, as in eq. 1, but only for the selected action.

$$\Delta \mathbf{W_A} = \alpha \delta X \bullet A'$$

Here $A'$ indicates a vector that is zero everywhere except the index corresponding to the selected action, which is equal to 1.

Besides the parameters governing the Gaussian activation function for each unit, the model was parameterized with a spatial discount factor ($\gamma = 0.8$) a softmax decision gain parameter ($\beta = 2$), and learning rates for value association, action associations, SRU position, and SRU spread ($\alpha_W = 0.01$, $\alpha_A = 0.01$, $\alpha_\mu = 0.1$ $\alpha_\sigma = 0.1$, respectively).

The agent's environment was a 9x9 grid world. Three different conditions were simulated in which the location of a reward source was placed in a corner, the center, or a center edge of the environment. We additionally simulated 2 conditions in which the initial distribution of the units was manipulated. In the uniform condition (Fig. 4A), a total of 25 SRUs were distributed to uniformly cover a grid extending 1 cell past the 9x9 grid world on each side. In the random condition, the 25 units were distributed at random over the same area. In all cases the initial $\sigma$ values were set to 0.75.

Trials were initialized by placing the agent in a random location along the edge of the grid world, and continued until the agent reached the reward source (trials could thus last forever for an unlucky agent, but in practice this never happened). A learning run consisted of 1000 trials, and 100 learning runs were conducted for each condition. Figures display results averaged over the final model values for all 100 runs.

## 4.4 | Category Representation

Two versions of the classification task were run, a one-dimensional (1D) version and a two-dimensional (2D) version. As in our spatial representation simulations, an actor-critic model was used, and both the actor and critic components received input from SRUs whose activity was determined by the location of a sample passed through each SRU's gaus-sian activation function. The goal of the model was to learn the correct categorization for samples randomly selected from a given range.

In the 1D version, samples were drawn from a uniform distribution with a range from 1 to 9. A total of 9 representation units were used, and their position was randomly initialized to a location from 0 to 10 (i.e., slightly beyond the edge of the space of possible samples). For all units, the initial value of $\sigma$ was set to 1. Other parameters include the actor's softmax decision gain parameter ($\beta$ = 15), and learning rates for value association, SRU position, and SRU spread ($\alpha_W$ =0.001, $\alpha_\mu$ = 0.01 $\alpha_\sigma$ = 0.01, respectively). These values were selected to emphasize the influence of changes in SRU position.

In the 2D version, samples were drawn from a uniform square area with a range from 1 to 9 in both directions. A total of 81 representation units were used, and their position was randomly initialized to a location from 0 to 10 along both axes (i.e., slightly beyond the edge of the space of possible samples). For all units, the initial value of $\sigma$ was set to 1. Other parameters include the actor's softmax decision gain parameter ($\beta$ = 15), and learning rates for value association, SRU position, and SRU spread ($\alpha_W$ =0.001, $\alpha_\mu$ = 0.1 $\alpha_\sigma$ = 0.1, respectively). As in the 1D case, these values were selected to emphasize the influence of changes in SRU position.

In both the 1D and 2D versions a single learning run consisted of 10000 independent trials, and 100 independent learning runs were conducted. Figures are derived from the model state at the end of each learning run.

## 4.5  |  Reversed RPEs

The simulations above were repeated, but for half of all SRUs (rounded down), learning rules for updating $\mu$ and $\sigma$ used the negative of the RPE. All other aspects of the simulations were identical.